\documentstyle[preprint,amsbsy]{ptptex}

\makeatletter
\def\YGrule{0.4}   
\def\YGbox{6.5}    
\def\SymBoxes#1#2#3#4{\newdimen\un@t \un@t#3%
\raisebox{#1}{\rule{#2\un@t}{#4}\hskip-#2\un@t
\@tempdimb\un@t \advance\@tempdimb by-#4\@tempcntb#2\relax%
\@whilenum{\@tempcntb>0}\do{
\rule{#4}{\un@t}\hskip\@tempdimb \advance\@tempcntb by\m@ne}%
\hskip-#2\un@t \rule[\un@t]{#2\un@t}{#4}%
\rule[\un@t]{#4}{#4}\hskip-#4
\rule{#4}{\un@t}}\hskip-#4}                
\def\Young{\@ifnextchar[{\@Young}{\@Young[0]}}
\def\@Young[#1]#2{\newdimen\YG@unit \YG@unit\YGbox pt%
\newdimen\h@ight \h@ight#1\YG@unit \@tempcnta-1\relax
\@tfor\c@ount:=#2\do{\advance\@tempcnta by\@ne}
\@tempdima\@tempcnta\YG@unit%
\advance\h@ight by\@tempdima\relax     
\@tfor\c@ount:=#2\do{\SymBoxes{\h@ight}{\c@ount}{\YG@unit}{\YGrule pt}%
\@tempdima-\c@ount\YG@unit \hskip\@tempdima%
\advance \h@ight by -\YG@unit}         
\@tempdima\YG@unit \multiply\@tempdima by\@car#2\@nil %
\hskip\@tempdima}                      
\def\YoungTab{\@ifnextchar[{\@YoungIdx}{\@YoungIdx[0]}}
\def\@YoungIdx[#1]{\@ifnextchar[{\@iYoungIdx[#1]}{\@iYoungIdx[#1][\@empty]}}
\def\@iYoungIdx[#1][#2]#3{%
\newdimen\YG@unit \YG@unit\YGbox pt\newdimen\YG@rule \YG@rule \YGrule pt
\newcount\c@ount \c@ount\z@ \newdimen\skip@wd \unitlength\@ne pt
\newdimen\h@ight \h@ight#1\YG@unit \@tempcnta\m@ne\relax
\@tfor\d@um:=#3\do{\advance\@tempcnta by\@ne}
\@tempdima\@tempcnta\YG@unit%
\advance\h@ight by\@tempdima\relax
\@tfor\@idxlist:=#3\do{
\@tempcnta\z@\hskip.5\YG@rule\relax 
\@for\@idx:=\@idxlist\do{
\raisebox{\h@ight}{\makebox(\YGbox,\YGbox){#2$\@idx$}}
\advance\@tempcnta by\@ne}\hskip-.5\YG@rule%
\@tempdima-\@tempcnta\YG@unit \hskip\@tempdima%
\ifnum\c@ount=\z@ \skip@wd-\@tempdima\fi \relax
\SymBoxes{\h@ight}{\@tempcnta}{\YG@unit}{\YG@rule}%
\hskip\@tempdima \advance\h@ight by -\YG@unit
\advance\c@ount by\@ne}
\hskip\skip@wd}                      
\def\lsim{\mathrel{\mathpalette\@versim<}}
\def\gsim{\mathrel{\mathpalette\@versim>}}
\def\@versim#1#2{\vcenter{\offinterlineskip
        \ialign{$\m@th#1\hfil##\hfil$\crcr#2\crcr\sim\crcr } }}
\makeatother

\def\cspan{\rule[-2.2ex]{0pt}{5ex}}
\def\mspan{\rule[-1ex]{0pt}{3ex}}

\def\m1{{-1}}

\def\VEV#1{\left\langle#1\right\rangle}
\def\mbf{\boldsymbol}
\newcommand{\nn}{\nonumber\\}

\def\gam#1{\,{}^{#1}\kern-1pt\Gamma}

\def\dim{\mathop{\rm dim}}

\preprintnumber{
KUNS-1803\\ hep-ph/0209088}

\title{Higgs Doublets as Pseudo Nambu-Goldstone Bosons \\ 
in Supersymmetric $\mbf{E_6}$ Unification}

\author{
Masako {\sc Bando}$^{1,}$\footnote{E-mail: bando@aichi-u.ac.jp}
and 
Taichiro {\sc Kugo}\footnote{E-mail: kugo@gauge.scphys.kyoto-u.ac.jp}
 }

\inst{%
$^1$Physics Division, Aichi University, Aichi 470-0296, Japan
\\
Department of Physics, Kyoto University, Kyoto 606-8502, Japan
}

\recdate{
\today}

\abst{
The idea to have Higgs doublets as pseudo Nambu-Goldstone (PsNG)
multiplets is examined in the framework of supersymmetric $E_6$ unified
theory. We show that extra PsNG multiplets other than the expected
Higgs doublets necessarily appear in the $E_6$ case. If we demand that
the extra PsNG multiplets neither disturb the gauge coupling
unification nor make the color gauge coupling diverge before
unification occurs, only possibility for the extra PsNG is
${\bf10}+\overline{\bf10}$ of $SU(5)$. This is realized when the
symmetry breaking $E_6 \rightarrow SO(10)$ occurs in the $\phi({\bf
27})+\phi(\overline{\bf27})$ sector while $E_6 \rightarrow
SU(4)_C\times SU(2)_L\times U(1)\times U(1)$ in the $\Sigma({\bf 78})$
sector. The existence of ${\bf10}+\overline{\bf10}$ multiplets with
mass around 1 TeV is therefore a prediction of this $E_6$ PsNG
scenario.  Implication of their existence on the proton decay is also
discussed.}
\begin{document}
\maketitle

There are many attractive features of grand unified theories (GUT), 
such as gauge unification, miraculous anomaly cancellation within  a
family, charge quantization, etc.. 
In the present 
form of GUT, however, there are also many unsolved problems.  One of 
the most serious difficulties would be the so-called hierarchy problem; 
we need extremely light Higgs doublets which are responsible for breaking 
electroweak symmetry, and 
their masses should be kept light against radiative corrections.
The most attractive way to protect against such radiative correction is to 
introduce supersymmetry (SUSY), which is not yet confirmed by 
experiments; no superpartner has been observed. 
Another aspect of the hierarchy problem is 
the so-called doublet-triplet (DT) splitting problem. 
It is not yet made clear how we can naturally split 
only $SU(2)_L$ doublets from their GUT partner color triplet states. 
There have been many attempts to solve this 
problem;\cite{DTsplitting,maekawa,DW,BarrRaby} 
 missing partner mechanism, sliding
 singlet mechanizm, Dimopoulos-Wilczek
 mechanism, etc..
Among various approaches we concentrate in this paper 
our attention on the simplest 
idea which has long been investigated, namely  the idea that 
 Higgs doublets are
realized as pseudo Nambu Goldstone (PsNG) 
bosons.\cite{inouekakuto,Anselm,Berezhiani:bd,Barbieri:1993wz,Berezhiani:1995sb,DP} 
Actually in supersymmetric grand unified scenario, once  
PsNG multiplets appear, 
they are kept massless so far as SUSY remains unbroken 
because of nonrenormalization theorem.  
Usually light PsNG multiplets are not welcome since their 
additional contributions to RGE harm the gauge unification.
However if we can identify them as the usual Higgs doublets,
 we can make active use of such property of  PsNG modes 
in explaining the light Higgs doublets.  

This letter aims to examine this idea of PsNG in supersymmetric 
$E_6$  unified theories. 
The idea of PsNG has been first proposed 
by Inoue, Kakuto and Takano in 1986\cite{inouekakuto}  
adopting a global $SU(6)$  
whose subgroup $SU(5)$ is  gauged.
Later it  was made more realistic  by 
Barbieri, Dvali and Moretti\cite{Barbieri:1993wz} by taking local 
$SU(6)$ symmetry and utilizing two Higgs sectors possessing no cross
couplings. Dvali and Pokorski\cite{DP} pointed out that the anomalous
$U(1)_X$ 
symmetry 
can play a role in making two Higgs sectors separated 
from each other in the superpotential term.
An extension to $E_6$ gauge symmetry was considered in
Ref.~\citen{Berezhiani:1995sb} with a negative result.  

Consider a supersymmetric grand unified theory based on a gauge group $G$. 
Suppose that the theory possesses two `Higgs scalar fields', $\phi$ and 
$\Sigma$, each of which need not be of irreducible representation of $G$ 
so that they each may actually stand for a set of fields. 
The point is that we assume that they have no direct cross couplings 
in the superpotential, 
\begin{equation}
W = W_1(\phi)+W_2(\Sigma),
\end{equation}
so that the superpotential has an enhanced symmetry $G_\phi\times G_\Sigma$, 
invariance under separate rotations of $\phi$- and 
$\Sigma$-sectors. In principle $G_\phi$ and $G_\Sigma$ can be (accidentally) larger 
than the gauge group $G$, but here we assume that both are $G$; 
$G_\phi=G_\Sigma=G$. Suppose that $\phi$ and $\Sigma$ develop their vacuum 
expectation values (VEVs) $\langle\phi\rangle$ and $\langle\Sigma\rangle$ and the symmetries are 
broken into
\begin{eqnarray}
G_\phi=G \quad &\rightarrow& \quad H_\phi\qquad \quad \hbox{by} \qquad \langle\phi\rangle,\nn
G_\Sigma=G \quad &\rightarrow& \quad H_\Sigma\qquad \quad \hbox{by} \qquad \langle\Sigma\rangle.
\end{eqnarray}
Then, the Nambu-Goldstone (NG) multiplets corresponding to the cosets 
$G/H_\phi$ and $G/H_\Sigma$ appear from the $\phi$ and $\Sigma$ sectors, 
respectively. 
But the actual symmetry of the full system is only $G$ and it is broken to 
the intersection subgroup $H_\phi\cap H_\Sigma$, so that the true NG multiplets 
are only those of $G/(H_\phi\cap H_\Sigma)$. The other multiplets 
not contained in $G/(H_\phi\cap H_\Sigma)$ are therefore all {\em pseudo 
Nambu-Goldstone} (PsNG) multiplets, whose number is counted as
\footnote{
This counting corresponds to the so-called maximum realization 
case.\cite{susynonl}}
\begin{eqnarray}
\hbox{\# of PsNG multiplets} &=&
\dim[G/H_\phi]+\dim[G/H_\Sigma]-\dim[G/(H_\phi\cap H_\Sigma)] \nn
&=&\dim G + \dim[H_\phi\cap H_\Sigma] - \dim H_\phi- \dim H_\Sigma.
\end{eqnarray}

Before entering the main subjects, we here comment on the the fact that 
exactly the same contents of PsNG multiplets also appear under a slightly
different setup which was originally considered by K.~Inoue and A.~Kakuto 
and H.~Takano. The setup they considered is as follows: the gauge symmetry 
$G_{\rm local}$ of the system is $H_\Sigma$, and the superpotential of 
the Higgs fields $\phi$ of the system possesses a global 
symmetry $G_{\rm global}=G$ larger than the required local symmetry 
$G_{\rm local}$ and $\phi$ develops a VEV which retains 
only a symmetry $H_\phi$. 
We call this setup `global $G$ setup' while the above one our `local $G$ 
setup'.
Note that we can exchange $H_\phi$ and $H_\Sigma$ in this global $G$ setup 
since our local $G$ setup is symmetric under the exchange $H_\phi 
\leftrightarrow H_\Sigma$.

The reason why the same contents of PsNG multiplets appear in both 
setups is as follows: Suppose that the VEV $\langle\Sigma\rangle$ is much larger than 
the VEV $\langle\phi\rangle$ in our local $G$ setup. Then we can consider an effective 
theory at the energy scale lower than $\langle\Sigma\rangle$ but higher than 
$\langle\phi\rangle$. There the original local symmetry $G$ is already spontaneously 
broken to $H_\Sigma$ and the associated NG multiplets of 
$G/H_\Sigma$ are all absorbed in the $G$-gauge multiplet. The rest components 
of $\Sigma$ become massive of order $\langle\Sigma\rangle$ and decouple. Therefore the 
system at this stage is just the same as that of the global $G$ setup 
with Higgs fields $\phi$. Indeed the superpotential of $\phi$ retains the 
symmetry $G$ as a global symmetry while the local gauge symmetry of the 
system is only $H_\Sigma$.  This finishes the proof. In this proof we have 
assumed $\langle\Sigma\rangle\gg\langle\phi\rangle$. But the number counting of 
broken generators is clearly independent of such an ordering, so the proof 
is generally valid.


First let us use the following notation\cite{DP} for the generated 
NG multiplets according to the representations under the 
standard theory gauge symmetry $G_S=SU(3)_C\times SU(2)_L\times{U(1)_Y}$:
\begin{eqnarray}
    \hat Q_Y &=& (3,2)_Y+(\bar 3,2)_{-Y},  \\ 
    \hat T_Y &=& (3,1)_Y+(\bar 3,1)_{-Y},  \\ 
    \hat D_Y &=& (1,2)_Y+(1,2)_{-Y} = \hat D_{-Y},  \\ 
         S_Y &=& (1,1)_Y.  
\end{eqnarray}
where the two numbers in each bracket stand for the dimensions of the 
representations of $SU(3)_C$ and $SU(2)_L$, and the attached 
suffix for the value
of the hypercharge $Y$. We will also use notation like $\hat Q$ 
when we do not specify the hypercharge value.

First of all let us find the representations of the true NG 
multiplets which appear when the group $E_6$ breaks down to the 
standard theory gauge group $G_S=SU(3)_C\times SU(2)_L\times U(1)_Y$.
The adjoint representation ${\bf 78}$ of $E_6$ is 
decomposed into irreducible representations of the subgroup $SO(10)$ 
as
\begin{equation}
{\bf78} = {\bf45}  + {\bf1}+ {\bf 16}+ \overline{\bf16},  
\end{equation}
and the $SO(10)$ adjoint ${\bf45}$ and the spinor ${\bf16}$ are 
further decomposed into $SU(5)$ representations as
\begin{eqnarray}
{\bf45} &=& {\bf24}  + {\bf1}+ {\bf 10}+ \overline{\bf10}, \nn
{\bf16} &=& {\bf10}  + \overline{\bf5}+ {\bf 1}.
\end{eqnarray}
As is well-known, these $SU(5)$ representations ${\bf24}$, ${\bf10}$ and
$\overline{\bf5}$ are decomposed under the standard theory gauge symmetry $G_S$ 
as\cite{ref:BK}
\begin{eqnarray}
{\bf24}&=& ({\bf8},{\bf1})_0 +({\bf1},{\bf3})_0
 +({\bf1},{\bf1})_0
+({\bf3},{\bf2})_{-5/6}
+(\overline{\bf3},{\bf2})_{5/6}, \nn
{\bf10}&=& ({\bf3},{\bf2})_{1/6} +(\overline{\bf3},{\bf1})_{-2/3}
 +({\bf1},{\bf1})_{-1}, \nn
\overline{\bf5}&=& (\overline{\bf3},{\bf1})_{1/3}
 +({\bf1},{\bf2})_{-1/2}.
\end{eqnarray}
Therefore, when $E_6$ breaks down to $SO(10)$, 
the NG multiplets appearing  
are given by
\begin{eqnarray}
E_6\rightarrow SO(10):\,\, {\bf 16}+ \overline{\bf16} +{\bf1}
&=&({\bf 10}+\overline{\bf10})+(\overline{\bf5}+{\bf5})+3\times{\bf1},  \nn
=(\hat Q_{1/6}&+&\hat T_{2/3}+S_1+S_{-1})
+(\hat T_{-1/3}+\hat D_{1/2})+ 3S_0,
\label{eq:SO10PsNG}
\end{eqnarray}
and, when $SO(10)$ further breaks down to $SU(5)$ and then to 
the standard theory gauge group $G_S$, the appearing NG multiplets are
\begin{eqnarray}
SO(10)\rightarrow SU(5)&:&\,\,{\bf 10}+ \overline{\bf10} +{\bf1} =
(\hat Q_{1/6}+\hat T_{2/3}+S_1+S_{-1}) +S_0, \nn
SU(5)\rightarrow G_S&:& \,\, ({\bf3},{\bf2})_{-5/6}
+(\overline{\bf3},{\bf2})_{5/6}
= \hat Q_{-5/6}.
\end{eqnarray}
The net NG multiplets appearing 
when $E_6$ breaks down to the standard theory gauge group $G_S$ is thus found 
to be 
\begin{eqnarray}
2(\hat Q_{1/6}+\hat T_{2/3}+S_1+S_{-1})
+\hat Q_{-5/6}
+(\hat T_{-1/3}+\hat D_{1/2})+ 4S_0.
\label{eq:trueNG}
\end{eqnarray}

Next, as another breaking pattern, we consider the breaking of $E_6$ 
into its maximal subgroup $SU(6)\times SU(2)$. The adjoint {\bf78} decomposes
under $SU(6)\times SU(2)$ as
\begin{equation}
{\bf78}=({\bf1},{\bf3}) + ({\bf35},{\bf1})+ ({\bf20},{\bf2}),
\label{eq:SU6*SU2}
\end{equation}
where the $SU(6)$ ${\bf20}$ of broken generator $({\bf20},{\bf2})$ is 
further decomposed under the subgroup $SU(4)\times SU(2)\subset SU(6)$
into
\begin{equation}
{\bf 20}= ({\bf4},{\bf1})+(\overline{\bf4},{\bf1}) +({\bf6},{\bf2})
\qquad \left(\leftarrow\ \ 
\YoungTab[-1]{{ }{ }{ }}=
\YoungTab[-1]{{ }{\cdot}{\cdot}}+
\YoungTab[-1]{{ }{ }{ }}+
\YoungTab[-1]{{ }{ }{\cdot}}
\ \right).
\label{eq:SU620}
\end{equation}
(The undotted and dotted boxes in the Young tableau on the 
right-hand side stand for the indices of $SU(4)$ and $SU(2)$ of 
the subgroup $SU(4)\times SU(2)\subset SU(6)$, respectively.)  
If the first factor group $SU(6)$ 
contains both $SU(3)_C$ and $SU(2)_L$ of the standard theory gauge group
$G_S$, in which case $SU(6)$ is denoted as $SU(6)_{C,L}$, the NG 
multiplets associate with the breaking $E_6\rightarrow SU(6)_{C,L}\times SU(2)$ are 
given by 
\begin{equation}
2\times{\bf20} =
2\times\left\{
(({\bf3},{\bf1})+(\overline{\bf3},{\bf1})) +2({\bf1},{\bf1}) +
(({\bf3},{\bf2})+(\overline{\bf3},{\bf2})) \right\} 
=2\hat Q + 2\hat T + 4S.
\end{equation}
Here we have not specified the hypercharge values since there are various 
possibilities how $U(1)_Y$ generators are embedded in the unbroken 
subgroup.
On the other hand, if $SU(3)_C$ is contained in the first $SU(6)$ while 
$SU(2)_L$ in the second $SU(2)$, i.e., 
$E_6$ breaks down to $SU(6)_C\times SU(2)_L$, then the resultant
NG multiplets are given by
\begin{equation}
({\bf20},{\bf2})= 
3\times({\bf3},\,{\bf2})
+3\times(\overline{\bf3},\,{\bf2})
+2\times({\bf1},\,{\bf2})
=3\hat Q + \hat D.
\end{equation}

Now let us consider the breaking patterns of $E_6$ into subgroups 
$H$ where $H$ contains the 
standard theory gauge group $G_S=SU(3)_C\times SU(2)_L\times U(1)_Y$.
In order to exhaust all the possibilities of the breaking patterns 
$E_6\rightarrow H$ in a systematic way,
we first classify the cases by identifying only the part $\tilde H$ of 
the subgroup $H$ 
containing the $SU(3)_C$ and $SU(2)_L$ groups of $G_S$. That is, we 
do {\em not} identify how the hypercharge $U(1)_Y$ is contained in the 
full $H$ and {\em neglect} the part (factor group) of $H$ which contains 
neither $SU(3)_C$ nor $SU(2)_L$. 
For instance, the choices of $H=SU(4)_C\times SU(2)_L\times SU(2)\times U(1)$ and  
$H=SU(4)_C\times SU(2)_L\times[U(1)]^k$ ($k=0,1,2$) are all classified into 
the case $\tilde H=SU(4)_C\times SU(2)_L$.
The suffices $C$ and $L$ attached to the group name always mean that 
the $SU(3)_C$ and $SU(2)_L$ groups of $G_S$ are contained in that group,
as we have defined in the above. This greatly simplify the task.

We classify the possibilities of the choice of $\tilde H$ according to its 
rank. The maximal regular subgroups of $E_6$ are 
$SU(6)\times SU(2)$, $SO(10)\times U(1)$ and $[SU(3)]^3$. However, since we only 
specify the factor groups that contain $SU(3)_C$ and $SU(2)_L$, then 
only possibilities of $\tilde H$ are clearly $SU(6)_C\times SU(2)_L$ and 
$SU(6)_{C,L}$ for the first $SU(6)\times SU(2)$, $SO(10)_{C,L}$ for the 
second $SO(10)\times U(1)$, and $SU(3)_C\times SU(3)_L$ for the third $[SU(3)]^3$.
Lower rank cases of $\tilde H$ can be found by considering further 
breaking of these cases. In this way we find all the possibilities 
for $\tilde H\supset G_S$ and tabulate them in Table.~\ref{table:1}. 
There we also list the representations of the NG multiplets under 
$\tilde H$ appearing in each breaking $E_6\rightarrow\tilde H$. 

\begin{table}[htb]
\caption{Possible choices for $\tilde H\supset G_S$ 
and NG fields for the breaking 
$E_6\rightarrow\tilde H$. The columns $\hat Q$, $\hat T$ and $\hat D$ denote the 
numbers of times those representations of NG multiplets appear 
in $E_6/\tilde H$. $SU(3)_C\times SU(2)_L$ singlets are neglected.}
\label{table:1}
\begin{center}
\begin{tabular}{|c|c|c|c|c|c|c|c|} \hline\hline
{\rm rank}   &Name  & $\tilde H$ 
& repr. under $\tilde H$ of the coset $E_6/\tilde H$
    &$\hat Q$    &$\hat T$   &$\hat D$   \\ \hline \hline 
6 & E &$SU(6)_C\times SU(2)_L$
& $({\bf 20,2})$ 
    &3          &0            &1        \\ \hline \hline 
5 &A  &$SO(10)_{C,L}$
& ${\bf16}+\overline{\bf16}$
    &1          &2          &1          \\ \hline 
&B  &$SU(5)_C\times SU(2)_L$ 
&  $({\bf10,2})+(\overline{\bf10},{\bf2})+({\bf5,1})
+(\overline{\bf5},{\bf1})$ 
    &3          &1            &1         \\ \hline
  &  &$SU(6)_{C,L}$ 
&$2\times{\bf 20}$
    &2           &2          &0         \\ \hline \hline 
4 &C  &$SU(5)_{C,L}$
&$2\times({\bf10}+\overline{\bf10})+{\bf5}+\overline{\bf5}$
    &2          &3           &1         \\ \hline
  &D  &$SU(4)_C\times SU(2)_L$ 
& $2\times\bigl(({\bf6,2})+({\bf4,1})+(\overline{\bf4},{\bf1})\bigr)
+({\bf4,2})+(\overline{\bf4},{\bf2})$
    &3           &2          &1        \\ \hline 
  &  &$SU(3)_C\times SU(3)_L$ 
&$3\times({\bf3,3})+3\times(\overline{\bf3},\overline{\bf3})$
    &3         &3             &0        \\ \hline  \hline 
3 &final    &$SU(3)_C\times SU(2)_L$ 
&$3\times 
({\bf3,2+1})+
3\times(\overline{\bf3},{\bf2+1})
+2\times({\bf 1,2})$
    &3           &3          &1      \\ \hline \hline  
\end{tabular}
\end{center}
\end{table}

Since we specify how $\tilde H$ contains $SU(3)_C$ and $SU(2)_L$, 
we can count the numbers of appearing NG multiplets 
of representations $\hat Q$, $\hat T$ and $\hat D$, which are also 
shown in Table.~\ref{table:1}. We can not count the numbers of 
$SU(3)_C\times SU(2)_L$-singlet NG multiplets nor the hypercharges of 
the $SU(3)_C\times SU(2)_L$ non-singlet NG multiplets. They can be 
specified later in concrete cases after narrowing down the possibilities.

Now with Table.~\ref{table:1}, we can find all the possible choices 
of $\tilde H_\phi$ and $\tilde H_\Sigma$. 
The conditions which should be satisfied are: 
i) an $SU(2)_L$ doublet $\hat D$ appear as a PsNG multiplet, and 
ii) other PsNG multiplets, if exist, should fall into an 
$SU(5)_{\rm GG}$ multiplet so as not to disturb the gauge coupling 
unification. 

{}From Table.~\ref{table:1}, we see that at most only one 
$\hat D$ NG multiplet can appear for any choices of $\tilde H$ and 
one $\hat D$ appears as a true NG multiplet in the $E_6\rightarrow G_S$ breakdown. 
In order to satisfy the condition i), therefore, 
we must have one $\hat D$ NG multiplet for each of breakings 
$E_6\rightarrow H_\phi$ and $E_6\rightarrow H_\Sigma$, and so the candidates for 
$\tilde H_\phi$ and $\tilde H_\Sigma$ are restricted to the cases 
A, B, C, D and E. 

For any choice of a pair $(H_\phi,\,H_\Sigma)$ from 
A, B, C, D and E, we immediately see that 
{\em extra PsNG multiplets appear} other than the desired 
$\hat D$ in this $E_6$ case. Note that the sum of the numbers of 
appearing $\hat Q$ and $\hat T$ in the pair should be larger than 
or equal to three for both $\hat Q$ and $\hat T$ 
since the true NG multiplets are 
$3\hat Q+3\hat T+\hat D$. If the sum is less than 3 for either 
$\hat Q$ or $\hat T$, it implies that 
the intersection $H_\phi\cap H_\Sigma$ is larger than $G_S$ in contradiction 
to the assumption. Since no extra $\hat D$ other than the two (a true NG
and a PsNG) multiplets appears, the only possibility for the $SU(5)$ 
multiplet into which other PsNG multiplets could fall is 
${\bf10}+\overline{\bf10}\supset 
\hat Q+\hat T$, which contains no $\hat D$ and 
equal numbers of $\hat Q$ and $\hat T$. Therefore the sums of 
the numbers of appearing $\hat Q$ and $\hat T$ 
should be equal in order to satisfy the condition ii). 

It is immediate to see that the only possible choices of such a pair 
satisfying this condition are (A,D) and (C,D). The former choice 
(A,D) yields  $4\hat Q+4\hat T+2\hat D$ so that it gives a 
${\bf10}+\overline{\bf10}$ extra PsNG multiplets, while 
the latter case (C,D) gives $5\hat Q+5\hat T+2\hat D$ containing 
two pairs of ${\bf10}+\overline{\bf10}$ extra PsNG multiplets. However 
we can see that the presence of $2\hat Q+2\hat T$ PsNG multiplets makes 
the $SU(3)_C$ gauge interaction asymptotically non-free and 
the coupling constant becomes infinity before reaching the unification 
scale. Indeed, we have the formula for the running coupling $\alpha=g^2/4\pi$ 
at one loop, 
\begin{eqnarray}
{1\over\alpha(\mu)}&=&{1\over\alpha(M)}+{b\over2\pi}\ln({M\over\mu}), \nn
b&=&-{9\over3}T({\rm adj})
+ \sum_RN_RT(R) \qquad {\rm tr}(T^a_RT^b_R)=T(R)\delta_{ab}
\end{eqnarray}
where $N_R$ is the number of chiral multiplets of representation $R$,
and the quadratic Casimir $T({\rm adj})\equiv C_2(G)$ is $N$ for $G=SU(N)$ and 
$T(\Young{1})=1/2$ for the fundamental representation $\Young{1}$ and  
$T(\Young[-.5]{11})=(N-2)/2$ for the representation $\Young[-.5]{11}$\,. 
For $SU(3)_C$ gauge coupling and for 
three generations (6 ${\bf3}+\overline{\bf3}$ chiral multiplets) plus 
two ${\bf10}+\overline{\bf10}$ PsNG multiplets 
($2\times(2+1)=6$ ${\bf3}+\overline{\bf3}$ chiral multiplets)), we have 
$b=-9+(6+6)(1/2+1/2)=3>0$, which makes $\alpha_s(\mu)$ diverge at around 
$\mu=6\times10^9$GeV. 
We thus see that the only possibility is the choice (A,D). It is 
interesting that the presence of $\hat Q+\hat T$ in this case just makes
the $\beta$ function of $SU(3)_C$ gauge coupling vanish at one-loop; 
$b=-9+(6+3)(1/2+1/2)=0$.

We thus have seen that the breaking pattern choice (A,D) is the only 
possibility. However, this is only a necessary condition. It is quite 
non-trivial whether there is actually a concrete model of breaking 
pattern (A,D) which also satisfies the $U(1)_Y$ quantum number 
requirements, which we have not examined above.

It is sufficient to find a model that satisfies all the 
requirements. We consider a model in which $E_6$ is spontaneously 
broken to $SO(10)_{C,L}$ by fundamental and anti-fundamental repr.~Higgs 
fields $\phi({\bf27})$ and $\phi(\overline{\bf27})$,
while it is broken 
down to $SU(4)_C\times SU(2)_L\times U(1)_A\times U(1)_B$ 
by an adjoint Higgs $\Sigma({\bf78})$:
\begin{eqnarray}
   {\rm A}\ :\quad    E_6   &\rightarrow& H_{\phi}=SO(10)_{C,L}\qquad 
\hbox{by}\quad \phi({\bf27})\ \hbox{and}\ \phi(\overline{\bf27}),  
\label{eq:Abreak}\\
   {\rm D}\ :\quad   E_6   &\rightarrow& H_{\Sigma}= SU(4)_C\times SU(2)_L\times U(1)_A\times U(1)_B 
        \qquad \hbox{by}\qquad \Sigma({\bf78}).
\label{eq:Dbreak}
\end{eqnarray}
(It should be noted that the breaking by adjoint $\Sigma$ cannot lower the 
rank of $H_\Sigma$ than that of $E_6$.) 
We shall specify these $SO(10)_{C,L}$, $SU(4)_C$ and $U(1)_A\times U(1)_B$ 
in more detail below by identifying which components of $\phi({\bf27})$ and 
$\Sigma({\bf78})$ acquire the VEVs. The requirements is that the 
intersection $H_{\phi}\cap H_{\Sigma}$ should be the standard model group $G_S$.

For that purpose, it is convenient to name all the twenty seven components 
of the fundamental representation $\phi({\bf27})$.
${\bf 27}$ is decomposed as ${\bf27}={\bf16}+{\bf10}+{\bf1}$ under 
Georgi-Fritsch-Minkowski's $SO(10)_{\rm GFM}\subset E_6$. 
Decomposing them further under 
Georgi-Glashow's $SU(5)_{\rm GG}\subset SO(10)$, 
we name the 27 components as follows:\cite{bk}
\begin{equation}
\begin{array}{ccccccccc}
 {\bf 16} & = &  {\bf 10} &+& {\bf 5}^* &+& {\bf 1}\ ,&&    \\
   &  &  \left[u^{ci}, \pmatrix{u_i\cr d_i\cr}, e^c\right] &&
(d^{ci}, e, -\nu) && \nu^c  &&     
\\[2ex]
 {\bf 10} & = &  {\bf 5} &+& {\bf 5}^*\ , 
&&  {\bf 1} & = &  {\bf 1}\ .   \\
      &  &  (D_i, E^c, -N^c ) && 
(D^{ci}, E, -N)  &&&& S\ \ \ \\
\end{array}
\label{eq:su510}
\end{equation}
The simplest scenario for the breaking A is realized by the VEV of 
the $SO(10)$-singlet 
component $S$ of $\phi({\bf 27})$:
\begin{equation}
\VEV{\mspan \phi({\bf1})=S(\phi)}=v_\phi. 
\end{equation}
In this case the unbroken subgroup $H_\phi$ is Georgi-Fritsch-Minkowski's 
$SO(10)_{\rm GFM}$ which contains Pati-Salam $SU(4)_{\rm PS}\simeq SO(6)$ 
and $SU(2)_L\times SU(2)_R\simeq SO(4)$ as its subgroup. 
But the choice of $SO(10)$ in $E_6$ even with 
a constraint $SO(10)\supset SU(5)_{\rm GG}$ is not unique at all but has a 
freedom of an $SU(2)$ rotation.
Indeed as pointed out in Ref.~\citen{ref:BK}, there is a maximal subgroup 
$SU(6)\times SU(2)_E$ in $E_6$, where $SU(6)\supset SU(5)_{\rm GG}$ and the 
$(5+1)\times2$ components in {\bf27}
\begin{equation}
\pmatrix{d^{c\,i} & e &  -\nu& -S \cr 
D^{c\,i} & E  & -N  &  -\nu^c\cr}\ 
\matrix{\leftarrow\ E_3=+1/2 \cr
\leftarrow\ E_3=-1/2 \cr}.
\end{equation}
give an $SU(2)_E$ doublet of $SU(6)$ {\bf6}-plets. That is, the two 
${\bf5}^*$-plets and two singlets ${\bf1}$ of $SU(5)_{\rm GG}$ 
in Eq.~(\ref{eq:su510}) are rotated into each other under the 
$SU(2)_E$. Since the generators of $SU(2)_E$ are orthogonal to those of 
$SU(5)_{\rm GG}$, the $SU(2)_E$-rotated $SO(10)$ from $SO(10)_{\rm GFM}$ 
with any angle $\mbf\theta$\cite{ref:BK}
\begin{equation}
SO(10)_{\mbf\theta}\equiv 
e^{i{\mbf\theta}\cdot{\mbf E}}SO(10)_{\rm GFM}
e^{-i{\mbf\theta}\cdot{\mbf E}}
\label{eq:twistedSO10}
\end{equation}
contains $SU(5)_{\rm GG}$ as its subgroup. 
Thus the VEV 
\begin{equation}
\cases{\VEV{S}=v_\phi\cos(\theta/2) \cr \VEV{\nu^c}=v_\phi\sin(\theta/2)\cr}
\qquad \leftrightarrow \qquad 
\left\{
\begin{array}{c}
\VEV{\mspan S_{\theta}=S\cos(\theta/2)+\nu^c\sin(\theta/2)}=v_\phi\\[1ex]
\VEV{\mspan \nu_\theta^c=\nu^c\cos(\theta/2)-S\sin(\theta/2)}=0
\end{array}\right.
\end{equation}
breaks $E_6$ down to a twisted $SO(10)$,  
$H_\phi=SO(10)_{\mbf\theta}$ (\ref{eq:twistedSO10}) 
with ${\mbf\theta}=(0,\theta,0)$. 
As a matter of fact, however, there is no loss of generality at 
this stage even if we assume that the $H_\phi$ symmetry is $SO(10)_{\rm 
GFM}=SO(10)_{{\mbf\theta}={\bf0}}$ with ${\mbf\theta}$ set equal to zero. 
This is because we have no reference frame at this stage and 
we are free to define those $SU(2)_E$-rotated fields 
$S_\theta$ and $\nu^c_\theta$ simply to be $S$ and $\nu^c$.  We can thus call 
$SO(10)_{\mbf\theta}$  simply $SO(10)_{\rm GFM}$.
If we have another reference frame, such as another VEV than $\VEV{\phi}$, 
then, this freedom of 
twisting $SO(10)$ becomes to have a physical meaning 
and we will actually use it below.

Next consider the D breaking (\ref{eq:Dbreak}) 
by the adjoint Higgs $\Sigma({\bf78})$. 
In order to specify 
the $SU(4)_C$ and $U(1)_A\times U(1)_B$ in the breaking pattern D, 
it is convenient to consider a maximal subgroup $SU(6)_C\times SU(2)_L$ in 
$E_6$, under which the fundamental {\bf27} decomposes into
\begin{equation}
(\overline{\bf15},{\bf1}) =
\pmatrix{\cspan -\varepsilon_{ikj}D^k & -u_i^c & -d_i^c & -D_i^c  \cr
                    u_j^c    &   0    &   S    & \nu^c    \cr
                    d_j^c    &  -S    &   0    & e^c     \cr
                    D_j^c    &  -\nu^c &   -e^c &  0      \cr},\quad
({\bf 6},{\bf 2}) =
\pmatrix{ \cspan u^i    & d^i   \cr
                  E^c   & - N^c \cr 
                  N     & E     \cr
                  \nu   & e     \cr}\ .
\label{eq:su6comp}
\end{equation}
Here the fist three entries and the last three entries of the 
{\bf6} of $SU(6)_C$ are the fundamental representations {\bf3} of 
$SU(3)_C$ and {\bf3} of $SU(3)_R$, respectively. The three components of 
{\bf3} of $SU(3)_R$ are arranged in the order for later convenience. 
We define and name three $SU(2)$ subgroups of the $SU(3)_R$ as 
follows by identifying their doublets:
\begin{equation}
 SU(2)_R:\pmatrix{ E^c   & - N^c \cr 
                   N     & E \cr},  \quad 
 SU(2)_{R'}:\pmatrix{ E^c   & - N^c \cr
                   \nu  &   e   \cr}, \quad
 SU(2)_E: \pmatrix{N    & E  \cr
                   \nu  & e  \cr}\ .
\label{eq:su2choice}
\end{equation}
The $SU(4)_C$ in the D breaking (\ref{eq:Dbreak}) should be 
$SU(4)_{C,\perp E}$ orthogonal to the $SU(2)_E$, 
whose fundamental representation {\bf4} is given by the 
first four entries in the $SU(6)$ representation (\ref{eq:su6comp}).
The reason is as follows.  

The true NG multiplets for the breaking $E_6\rightarrow G_S$ are given in 
Eq.~(\ref{eq:trueNG})
\begin{equation}
2(\hat Q_{1/6}+\hat T_{2/3}+S_1+S_{-1})
+\hat Q_{-5/6}
+(\hat T_{-1/3}+\hat D_{1/2})+ 4S_0.
\label{eq:trueNG2}
\end{equation}
In addition to these we expect in this (A,D) breaking scenario
that there appear the following PsNG multiplets:
\begin{equation}
({\bf 10}+\overline{\bf10})+\hat D_{1/2} +x\,S_0,
=(\hat Q_{1/6}+\hat T_{2/3}+S_1+S_{-1})+\hat D_{1/2} +x\,S_0,
\end{equation}
where the number $x$ of $G_S$-singlets $S_0$ can be arbitrary. 
On the other hand, the NG multiplets coming from the $\phi$-sector 
in which $E_6\rightarrow SO(10)_{\rm GFM}$ occurs are given in Eq.~(\ref{eq:SO10PsNG}):
\begin{eqnarray}
(\hat Q_{1/6}+\hat T_{2/3}+S_1+S_{-1})
+(\hat T_{-1/3}+\hat D_{1/2})+ 3S_0.
\end{eqnarray}
Therefore the NG multiplets appearing from the $\Sigma$-sector should be
\begin{equation}
2(\hat Q_{1/6}+\hat T_{2/3}+S_1+S_{-1})
+\hat Q_{-5/6}
+\hat D_{1/2}+ (1+x)S_0.
\label{eq:SigmaNG}
\end{equation}
Note that the breaking in the $\Sigma$-sector is 
$E_6\rightarrow SU(4)_C\times SU(2)_L\times U(1)_A\times U(1)_B$ while the eventual breaking  
accompanied by the true NG multiplets is 
$E_6\rightarrow G_S=SU(3)_C\times SU(2)_L\times U(1)_Y$. So the difference 
between 
(\ref{eq:SigmaNG}) and (\ref{eq:trueNG2}),
\begin{equation}
\hat T_{-1/3}+(3-x)S_0
\end{equation}
must correspond to the NG multiplets associated with 
the breaking 
$SU(4)_C\times U(1)_A\times U(1)_B\rightarrow SU(3)_C\times U(1)_Y$.
For the latter breaking we generally have $\hat T_Y+2S_0$ as NG multiplets 
(so that $x$ is fixed to be 1). In order for this color triplet 
$\hat T_Y$ for the breaking 
$SU(4)_C\rightarrow SU(3)_C$ to carry the desired hypercharge $Y=-1/3$, the 
difference of $Y$ quantum number of the first three color triplet 
components from that of the fourth component of $SU(4)_C$ {\bf4} 
should be $-1/3$.
Noting the hypercharge quantum numbers 
$Y((u^i,d^i))=1/6$, $Y((E^c,-N^c))=+1/2$,
$Y((N,E))=-1/2$ and $Y((\nu,e))=-1/2$, we see that the only possibility 
for $SU(4)_C$ is $SU(4)_{C,\perp E}$ for which the {\bf4} is given by
\begin{equation}
({\bf 6},{\bf 2}) =
\pmatrix{ \cspan u^i    & d^i   \cr
                  E^c   & - N^c \cr} .
\end{equation}
Indeed then the generator which converts the fourth entry $E^c$ to 
$u$-quark $u^i$ is $SU(3)_C$ color triplet {\bf3} and carries hypercharge 
$Y(u^i)-Y(E^c)=1/6-1/2=-1/3$ as required.

%

Now let us identify the VEV of $\Sigma({\bf78})$ which realizes such D 
breaking $E_6\rightarrow SU(4)_{C,\perp E}\times SU(2)_L\times U(1)_A\times U(1)_B$. As we have seen 
in Eq.~(\ref{eq:SU6*SU2}), the adjoint $\Sigma({\bf78})$ is decomposed under
$SU(6)_C\times SU(2)_L$ as ${\bf78}=({\bf1},{\bf3}) + ({\bf35},{\bf1})+ 
({\bf20},{\bf2})$, the VEV $\VEV{\Sigma}$ realizing such a breaking 
is developed in the $SU(6)_C$ adjoint component $({\bf35},{\bf1})$:
\begin{equation}
\VEV{\Sigma({\bf 35,1})}=
\pmatrix{ a{\bf1}_4 & 0 & 0 \cr
             0     & b & 0 \cr
             0     & 0 & c \cr} \qquad (4a+b+c=0)
\label{eq:sigmaVEV}
\end{equation}
Here this $6\times6$ matrix is written on the same basis as in 
Eq.~(\ref{eq:su6comp}) so that the bottom right $2\times2$ submatrix 
corresponds to $SU(2)_E\times U(1)$. Note that we have used $SU(6)_C$ rotations 
to bring the generic VEV of hermitian $6\times6$ matrix $\Sigma({\bf 35,1})$ 
into the above diagonal form; in particular, an $SU(2)_E$ rotation is 
used to make the bottom right $2\times2$ submatrix diagonal. This means that 
the previous $\phi$-sector unbroken subgroup $H_\phi$ 
no longer remains to be the $SO(10)_{\rm GFM}$ with 
$\theta=0$ in this basis but becomes $SO(10)_{\mbf\theta}$ with $\theta\not=0$. 
For $\theta\not=0$ to have a physical meaning, the $SU(2)_E$ must be broken by 
$b\not=a$ as we assume here. Then two unbroken $U(1)$ charges, called 
$U(1)_A$ and $U(1)_B$ in the above, are given in this basis by 
\begin{equation}
U(1)_A \ : \ \ A\equiv 
\pmatrix{ {\bf1}_4 & 0 & 0 \cr
             0     & -2 & 0 \cr
             0     & 0 & -2 \cr}, \qquad 
U(1)_B \ : \ \ B\equiv 
\pmatrix{ {\bf0}_4 & 0 & 0 \cr
             0     & 1 & 0 \cr
             0     & 0 & -1 \cr} = E_3.
\end{equation}
The latter charge $B$ is chosen to be the third component $E_3$ of 
$SU(2)_E$. 

It should be emphasized that $\theta$ must not be zero. Otherwise, the 
intersection $H_\phi\cap H_\Sigma$ would contain an extraneous $U(1)$ other than 
the standard theory gauge symmetry $G_S$. Indeed, if $H_\phi=SO(10)_{\rm 
GFM}$, its five Cartan generators are all diagonal in the particle basis
which we have defined in Eq.~(\ref{eq:su510}), 
while $H_\Sigma=SU(4)_{C,\perp E}\times SU(2)_L\times U(1)_A\times U(1)_B$ is rank 6 and 
contains all the Cartan generators in $E_6$, which are also diagonal on 
the same basis. Therefore the $U(1)_V$ contained in $SO(10)_{\rm GFM}\supset 
SU(5)_{\rm GG}\times U(1)_V$ can be necessarily written as a linear 
combination of the six Cartan generators in $H_\Sigma$ and hence remains as 
an unbroken symmetry contained in the intersection $H_\phi\cap H_\Sigma$ in 
contradiction to the assumption. 
If $\theta\not=0$, on the other hand, the 
directions of Cartan generators in $H_\phi$ and $H_\Sigma$ are twisted and no 
such $U(1)$ remains. [This can be seen by looking at, e.g., $e^{-i\theta 
E_2}E_3e^{i\theta E_2}= E_3\cos\theta+E_1\sin\theta$.]

Finally let us confirm the quantum numbers including the hypercharge of 
the NG multiplets which actually appear in this D breaking 
$E_6\rightarrow SU(4)_{C,\perp E}\times SU(2)_L\times U(1)_A\times U(1)_B$ realized by the $\Sigma$-VEV 
(\ref{eq:sigmaVEV}). Noting the hypercharge $Y$ is given by 
\begin{equation}
Y=\pmatrix{ 
(1/6){\bf1}_3 & 0 & 0 \cr
             0     & 1/2 & 0 \cr
             0     & 0 & (-1/2){\bf1}_2 \cr}, \qquad 
\end{equation}
on the fundamental representation {\bf6} of $SU(6)_C$ 
in the basis (\ref{eq:su6comp}), we can find the 
$SU(3)_C\times U(1)_Y$ quantum numbers of 
the ${\bf20}=\Young[-1]{111}$ by inspecting Eq.~(\ref{eq:SU620}):\footnote{%
\def\YGbox{8}
For instance, the hypercharge $Y=-5/6$ for $({\bf3}_{-5/6},{\bf1})$ 
can be found as follows. It corresponds to 
$\YoungTab[-1]{{\alpha}{\cdot}{\cdot}}$ with color index $\alpha=1,2,3$ and 
$\cdot=5$ or 6, hence carrying the hypercharge 
$(1/6)+(-1/2)+(-1/2)=-5/6$.}
\begin{eqnarray}
{\bf 20} &=& ({\bf4},{\bf1})+(\overline{\bf4},{\bf1}) +({\bf6},{\bf2}) 
\qquad \hbox{under}\ SU(4)_{C,\perp E}\times SU(2)_E \nn
&=& ({\bf3}_{-5/6}+{\bf1}_{-1/2},{\bf1})
+(\overline{\bf3}_{5/6}+{\bf1}_{1/2},{\bf1})
+({\bf3}_{1/6}+\overline{\bf3}_{-1/6},{\bf2}) 
\end{eqnarray}
so that the $G_S$ quantum numbers of $({\bf20,2})$,
which appears for the breaking $E_6\rightarrow SU(6)_C\times SU(2)_L$, are given by
\begin{equation}
({\bf20,2}) = \hat Q_{-5/6} + \hat D_{1/2} + 2\hat Q_{1/6}.
\label{eq:38}
\end{equation}
When $SU(6)_C\times SU(2)_L$ is further broken to 
$SU(4)_{C,\perp E}\times SU(2)_L\times U(1)_A\times U(1)_B$, the appearing NG multiplets are
\begin{eqnarray}
&&({\bf4},\overline{\bf2})+(\overline{\bf4},{\bf2}) + 2S_0
\qquad \hbox{under}\ SU(4)_{C,\perp E}\times SU(2)_E \nn
&&\quad = ({\bf3},\overline{\bf2})_{2/3}
+({\bf1},\overline{\bf2})_1
+(\overline{\bf3},{\bf2})_{-2/3}
+({\bf1},{\bf2})_{-1} +2S_0 \nn
&&\quad = 2\times(\hat T_{2/3}+S_1+S_{-1}) + 2S_0 
\label{eq:39}
\end{eqnarray} 
where $2S_0$ comes from the breaking $SU(2)_E\rightarrow U(1)_{E_3}$.
We thus see that the resultant NG multiplets (\ref{eq:38}) plus 
(\ref{eq:39}) 
indeed realizes the expected one in Eq.~(\ref{eq:SigmaNG}).

{\bf Proton Decay}: 

The important prediction of the present idea of Higgs doublets as PsNG 
multiplets is that there necessarily appear additional PsNG multiplets 
${\bf10}_H+\overline{\bf10}_H$ of $SU(5)_{\rm GG}$ which we expect will 
get masses $M_{\bf10}$ around $O(1)$TeV after SUSY is broken. Aside from
the direct observation of them, their effect may be seen through proton 
decay. Let us evaluate the order of the proton decay caused by their 
effect.

We expect generically the presence of the following dimension 4 and 5 
operators in the 
low energy effective superpotential: in terms of the $SU(5)$ language,
\begin{eqnarray}
W_4 &=& f_4^{ij}\overline{\bf5}_i\overline{\bf5}_j{\bf10}_H 
\supset f_4^{ij}[\epsilon^{\alpha\beta\gamma}d^c_{i\alpha}d^c_{j\beta}u^c_{H\gamma} 
+ d^c_{i\alpha}(e_ju_H^\alpha-\nu_jd_H^\alpha)] \nn
W_5 &=& {f_5^{ij}\over M_{\rm pl}}
{\bf10}_i\overline{\bf5}_j{\bf5}_H\overline{\bf10}_H 
\supset f_5^{ij}[\epsilon^{\alpha\beta\gamma}u^c_{i\alpha}d^c_{j\beta}
+ (u^\gamma_ie_j-d^\gamma_i\nu_j)]\VEV{H_u}\overline d_{H\gamma},
\end{eqnarray}
where ${\bf10}_i$ and $\overline{\bf5}_i$ ($i=1,2,3$) denotes three 
generations of matters, ${\bf10}_H$ and $\overline{\bf10}_H$ are our new
light Higgs, and ${\bf5}_H$ is the usual Higgs $H_u$ in which the color 
triplet part is in fact missing. If the colored components in 
${\bf10}_H$ and in $\overline{\bf10}_H$ are connected by propagator 
$\langle{\bf10}_H\overline{\bf10}_H\rangle$ and the usual Higgs doublet 
${\bf5}_H$ is replaced by the VEV $\VEV{H_u}$, then we have an 
effective superpotential which breaks baryon number:
\begin{equation}
W_6 = 
{f^{ijkl}_{\rm eff}\over M_{\rm pl}}
\epsilon^{\alpha\beta\gamma}u^c_{i\alpha}d^c_{j\beta}
\times d^c_{k\gamma}\nu_l\ , \qquad 
\quad 
f^{ijkl}_{\rm eff}=f_5^{ij}f_4^{kl} 
{\VEV{H_u}\over M_{\bf10}}
\end{equation}
Note that if the Higgs VEV $\VEV{H_u}$ is replaced by the Higgs 
superfield, then this term gives an dimension 6 operator but the 
suppression is not by the square of Planck mass $M_{\rm pl}$ 
but by a single power of $M_{\rm pl}$. Another mass scale $M_{\bf10}$ 
comes from 
the propagator of ${\bf10}_H$ Higgs which is light and does not give 
any significant suppression;
$\VEV{H_u}/M_{\bf10}\sim 1 - 10^{-1}$. 
So this operator is potentially dangerous so that the proton decay by this 
operator should be suppressed by the smallness of the coupling constant. 

Similarly to the analysis of the generic dimension 5 operators as performed by Kakizaki and Yamaguchi,\cite{kakizaki} 
we can think that the coupling constants $f_4^{ij}$ and $f_5^{ij}$ 
obey a Froggatt-Nielsen\cite{FN} suppression mechanism 
similar to the usual Higgs Yukawa coupling 
constant responsible for the fermion masses.  
Then, using letters $q_i, l_i, u_i^c, d_i^c$ and $h_u, h_d$ 
to denote the Froggatt-Nielsen $U(1)$ charges of $i$-th generation quarks 
and leptons and the up- and down-type Higgs doublets $H_u$ and $H_d$,
\begin{eqnarray}
f_{5}^{ij}f_4^{kl} &=& f_4f_5 \lambda^{u^c_i+d^c_j+d^c_k+l_l+h_u}, 
\qquad f_{4,5}\sim O(1) \nn
&=& (f_4f_5) y_d^{ij}\lambda^{(u^c_i-q_i)-h_d}y_d^{lk}\lambda^{(l_l-q_l)-h_d+h_u}, 
\end{eqnarray}
where $\lambda\sim\sin\theta_C=0.22$ and $y^{ij}_d=\lambda^{q_i+u^c_j+h_d}$ is the 
$ij$ matrix element of the down-type quark yukawa coupling.
Then the largest operator is 
\begin{equation}
W_6 = 
{f^{1123}_{\rm eff}\over M_{\rm pl}}u_Rd_Rs_R \nu_\tau\ , 
\end{equation}
so that the main decay mode is $p\rightarrow K^+\overline\nu_\tau$.
The bound for the proton lifetime $\tau_{\rm proton}>2\times10^{33}$yr 
gives a constraint
\begin{equation}
|f^{ijkl}_{\rm eff}| 
\lsim \left(10^{-7}\sim\lambda^{11}\right)
\times\left({M_{\rm pl}\over10^{19}{\rm GeV}}\right)
\end{equation}
We have for the main decay mode
\begin{equation}
f^{1123}_{\rm eff} \simeq 
(f_4f_5) {\VEV{H_u}\over M_{\bf10}}\times\left(
y_d^{11}\lambda^{(u^c_1-q_1)-h_d}y_d^{32}\lambda^{(l_3-q_3)-h_d+h_u}
= 
y_d^{11}y_d^{32}
\lambda^{p+2h_u-3h_d}\right)
\end{equation}
where use has been made of the `GUT-inspired' relations 
$q_i=u^c_i$ and $l_i=d_i^c$
by Kakizaki and Yamaguchi and of the definition of $p$:
\begin{equation}
y_b/y_t= \lambda^{d^c_3+h_d-u^c_3-h_u} \equiv\lambda^p 
\quad \rightarrow \quad l_3-q_3 = p +h_u-h_d
\end{equation}
If we use $p=2$ corresponding $\tan\beta\simeq 3$, and 
semi-empirical relations $y_d^{11}=\lambda^5y_b$, 
$y_d^{32}=y_d^{33}\equiv y_b$ and $y_t\sim1$,
we have
\begin{equation}
f^{1123}_{\rm eff} \simeq 
(f_4f_5) {\VEV{H_u}\over M_{\bf10}}\times 
\lambda^{11+2h_u-3h_d}
\end{equation}
Therefore, since it is natural to expect that 
the factors $(f_4f_5) {\VEV{H_u}\over M_{\bf10}}$ and $\lambda^{2h_u-3h_d}$ are 
of order 1, we could see the proton decay in near future. 

{\bf Yukawa couplings}: 

The particular property of our Higgs doublets as PsNG multiplets is 
their representations under $SO(10)\subset E_6$. As is seen from the 
discussion above, in particular Eq.~(\ref{eq:SO10PsNG}), the down-type 
Higgs $H_d$ is contained in $({\bf16},\overline{\bf5})$ in $\phi({\bf27})$ 
and up-type Higgs $H_u$ in $(\overline{\bf16},{\bf5})$ in 
$\phi(\overline{\bf27})$, where the two numbers in the brackets denote 
representations under $SO(10)$ and $SU(5)$.
This is in sharp contrast with the usual GUTs in which $H_u$ and $H_d$ 
are assigned to be $({\bf10},{\bf5})$ and $({\bf10},\overline{\bf5})$.
This property leads to some peculiarities in obtaining fermion mass 
terms in this model. 
The down-type quark mass terms come from the 
usual trilinear terms but they actually exist only when 
the down-type quarks contain `$SU(2)_E$ twisted components' 
$\Psi({\bf10},\overline{\bf5})$:
\begin{equation}
\Psi_i({\bf 27}) \Psi_j({\bf 27}) \phi({\bf27}) 
\  \rightarrow \ \Psi_i({\bf16,10}) \Psi_j
({\bf10},\overline{\bf5})\phi({\bf16},\overline{\bf5}).
\end{equation}
Since $\phi({\bf27})$ does not contain the up-type Higgs $H_u\subset{\bf5}$, 
these trilinear terms do not contain up-type quark masses at all.
Up-type quark mass terms come from dimension 5 operators:
\begin{equation}
\Psi_i({\bf 27})\Psi_j({\bf 27})\phi(\overline{\bf27})\phi(\overline{\bf27})
\ \rightarrow \ \Psi_i({\bf16,10}) \Psi_j({\bf16,10})
\phi(\overline{\bf16},\overline{\bf5})
\VEV{\phi(\overline{\bf16},\overline{\bf1})}.
\end{equation}
Note that the VEV $\langle\phi(\overline{\bf16},\overline{\bf1})\rangle$ 
is non-vanishing only when the $SU(2)_E$ rotation (\ref{eq:twistedSO10})
in the Higgs sector exists, $\theta\not=0$. 
The induced top Yukawa copling is thus not of dimension 4 coupling but 
comes from a  higher dimensional
operator. The resultant Yukawa couplings are thus accompanied with  
 $\langle\phi(\overline{\bf16},\overline{\bf1})\rangle/M_P$.
This eventually suppress the top Yukawa coupling by the power 
 $\lambda$ or so. Note that the bottom Yukawa coupling 
can in principle be  dimension 4. However we expect the 
so-called family twisting structure\cite{ref:BK,bm,bk} and so the bottom 
Yukawa couplings may be accompanied by some Froggatt-Nielsen 
factor, so that the ratio of Yukawa couplings of the top and the bottom quarks 
can become smaller. Also note that 
in our scenario the unified gauge coupling is larger than 
the usual case and it may be possible to get a reasonable top quark mass  
as an quasi infrared fixed point; the running Yukawa  coupling 
approaches to the order of color 
gauge coupling faster than in the usual case
\footnote{The same problem already happens in the $SU(6)$ 
case considered by 
Dvali and Pomarol.\cite{DP} They introduced an additional  
fermion fields  ${\bf 20}$, and top quark is represented as 
a mixture state of ${\bf 15}$ with  ${\bf 20}$. 
This provides a dimension 4 top Yukawa coupling.}


We conclude this note by adding some comments.

The PsNG Higgs approach based on the model with 
$G=SU(6)\times SU(2)_R$ gauge symmetry instead of $E_6$ 
may also be interesting,\cite{matsuoka} 
in which the breaking pattern is given by
\begin{eqnarray}
{}[SU(6)\times SU(2)_R]_{\phi} &\longrightarrow &SU(5)_{\rm GG},\\
{}[SU(6)\times SU(2)_R]_\Sigma&\longrightarrow &SU(4)_C\times SU(2)_L\times SU(2)_R\times U(1)
\end{eqnarray}
and there appears no extra PsNG multiplet than the desired Higgs 
doublets.
This breaking pattern can be realized by the 
$\phi$ and $\Sigma$ Higgs sectors which consist of 
$\phi(\overline{\bf6},{\bf2})$ and $\Sigma({\bf 15,1})$ 
in addition to their conjugates, respectively. 
Note that $\Sigma({\bf 15,1})$ contains no $SU(5)$ singlet component 
but has an $SU(4)$ singlet. 
So it can naturally breaks $SU(6)$ down to $SU(4)$ instead $SU(5)$. 
Moreover these Higgs fields $\phi(\overline{\bf6},{\bf2})$ 
and $\Sigma({\bf 15,1})$ can be combined into a 
fundamental representation ${\bf 27}$ of $E_6$ representation. 
So if $E_6$ is broken by some mechanism, for example by Hosotani 
mechanism, it may be possible to make a realistic scenario by 
using only the fundamental representation Higgs. 

The notion of PsNG bosons were first investigated intensively by S. 
Weinberg\cite{weinberg}
in the context of dynamical symmetry breaking. 
The essential difference between his PsNG and the present one is 
the existence of SUSY.  In the non-SUSY case 
the mass of PsNG is generated via residual 
gauge interaction which breaks the tree level symmetry 
and the order is estimated to be $m^2=g^2 \Lambda^2$, with 
$\Lambda$ being a characteristic scale of the interaction responsible for 
the spontaneous breaking. 
In SUSY case, on the other hand, the masses of PsNG fields are protected 
until the SUSY breaking occurs. This ensures masses of PsNG very light 
of the order $\sim gM_{\rm SUSY}$. 

We would like to stress that $E_6$\cite{ref:E6} 
model has many advantages. 
Especially after the recent neutrino oscillation observations 
 confirmed the remarkable fact of the neutrino large mixings, 
$E_6$ model became more attractive because we anyhow need some 
non-parallel (twisting) family structure in order to 
reproduce those large mixings.\cite{yanagida} 
$E_6$ provides us with the most natural scenario for realizing this 
twisting family structure.\cite{bk,bm} 
We have seen in this paper that this twisting structure is also 
required in the symmetry breaking pattern to assure the 
intersection $H_\phi\cap H_\Sigma$ reduces to the standard theory gauge group 
$G_S$.
\subsection*{Acknowledgments}

We would like to thank to H.~Haba, N.~Maekawa, S..~Yamashita, 
and M.~Kakizaki for stimulating discussions. 
M.~B.\ and T.~K.\ 
are supported in part by the Grant-in-Aid for Scientific Research 
Nos.~12640295 and 13640279, respectively, 
from Japan Society for the Promotion of Science, and 
Grants-in-Aid for Scientific Research on Priority Area A 
``Neutrinos" (Y.~Suzuki) Nos.~12047225 and 12047214, respectively,
from the Ministry of Education, Science, Sports and Culture, Japan.
Also we are stimulated by  the fruitful and instructive discussions 
during the Summer Institute 2001 and 2002 held at 
Fuji-Yoshida.

\end{document}